\documentstyle[twoside]{article}
 
\catcode`\@=11
\long\def\@makefntext#1{
\protect\noindent \hbox to 3.2pt {\hskip-.9pt
$^{{\eightrm\@thefnmark}}$\hfil}#1\hfill}               
 
\def\thefootnote{\fnsymbol{footnote}}
\def\@makefnmark{\hbox to 0pt{$^{\@thefnmark}$\hss}}    
 
\def\ps@myheadings{\let\@mkboth\@gobbletwo
\def\@oddhead{\hbox{}
\rightmark\hfil\eightrm\thepage}
\def\@oddfoot{}\def\@evenhead{\eightrm\thepage\hfil
\leftmark\hbox{}}\def\@evenfoot{}
\def\sectionmark##1{}\def\subsectionmark##1{}}
 
\oddsidemargin=\evensidemargin
\renewcommand{\thefootnote}{\fnsymbol{footnote}}

\newcounter{sectionc}\newcounter{subsectionc}\newcounter{subsubsectionc}
\renewcommand{\section}[1] {\vspace{12pt}\addtocounter{sectionc}{1}
\setcounter{subsectionc}{0}\setcounter{subsubsectionc}{0}\noindent
        {\tenbf\thesectionc. #1}\par\vspace{5pt}}
\renewcommand{\subsection}[1] {\vspace{12pt}\addtocounter{subsectionc}{1}
        \setcounter{subsubsectionc}{0}\noindent
        {\bf\thesectionc.\thesubsectionc. {\kern1pt \bfit #1}}\par\vspace{5pt}}
\renewcommand{\subsubsection}[1] {\vspace{12pt}\addtocounter{subsubsectionc}{1}
        \noindent{\tenrm\thesectionc.\thesubsectionc.\thesubsubsectionc.
        {\kern1pt \tenit #1}}\par\vspace{5pt}}
\newcommand{\nonumsection}[1] {\vspace{12pt}\noindent{\tenbf #1}
        \par\vspace{5pt}}
 
\newcounter{appendixc}
\newcounter{subappendixc}[appendixc]
\newcounter{subsubappendixc}[subappendixc]
\renewcommand{\thesubappendixc}{\Alph{appendixc}.\arabic{subappendixc}}
\renewcommand{\thesubsubappendixc}
        {\Alph{appendixc}.\arabic{subappendixc}.\arabic{subsubappendixc}}
 
\renewcommand{\appendix}[1] {\vspace{12pt}
        \refstepcounter{appendixc}
        \setcounter{figure}{0}
        \setcounter{table}{0}
        \setcounter{lemma}{0}
        \setcounter{theorem}{0}
        \setcounter{corollary}{0}
        \setcounter{definition}{0}
        \setcounter{equation}{0}
        \renewcommand{\thefigure}{\Alph{appendixc}.\arabic{figure}}
        \renewcommand{\thetable}{\Alph{appendixc}.\arabic{table}}
        \renewcommand{\theappendixc}{\Alph{appendixc}}
        \renewcommand{\thelemma}{\Alph{appendixc}.\arabic{lemma}}
        \renewcommand{\thetheorem}{\Alph{appendixc}.\arabic{theorem}}
        \renewcommand{\thedefinition}{\Alph{appendixc}.\arabic{definition}}
        \renewcommand{\thecorollary}{\Alph{appendixc}.\arabic{corollary}}
        \renewcommand{\theequation}{\Alph{appendixc}.\arabic{equation}}
        \noindent{\tenbf Appendix \theappendixc #1}\par\vspace{5pt}}
\newcommand{\subappendix}[1] {\vspace{12pt}
        \refstepcounter{subappendixc}
        \noindent{\bf Appendix \thesubappendixc. {\kern1pt \bfit #1}}
        \par\vspace{5pt}}
\newcommand{\subsubappendix}[1] {\vspace{12pt}
        \refstepcounter{subsubappendixc}
        \noindent{\rm Appendix \thesubsubappendixc. {\kern1pt \tenit #1}}
        \par\vspace{5pt}}
 
\newcommand{\textlineskip}{\baselineskip=13pt}
\newcommand{\smalllineskip}{\baselineskip=10pt}
 
\def\eightcirc{
\begin{picture}(0,0)
\put(4.4,1.8){\circle{6.5}}
\end{picture}}
\def\eightcopyright{\eightcirc\kern2.7pt\hbox{\eightrm c}}
 
\newcommand{\copyrightheading}[1]
        {\vspace*{-2.5cm}\smalllineskip{\flushleft
        {\footnotesize $\eightcopyright$\, World Scientific Publishing
         Company}\\
         }}
 
\def\abstracts#1#2#3{{
        \centering{\begin{minipage}{4.5in}\baselineskip=10pt\footnotesize
        \parindent=0pt #1\par
        \parindent=15pt #2\par
        \parindent=15pt #3
        \end{minipage}}\par}}
 
\renewenvironment{thebibliography}[1]
        {\frenchspacing
         \ninerm\baselineskip=11pt
         \begin{list}{\arabic{enumi}.}
        {\usecounter{enumi}\setlength{\parsep}{0pt}
         \setlength{\leftmargin 12.7pt}{\rightmargin 0pt} 
         \setlength{\itemsep}{0pt} \settowidth
        {\labelwidth}{#1.}\sloppy}}{\end{list}}
 
\newcounter{itemlistc}
\newcounter{romanlistc}
\newcounter{alphlistc}
\newcounter{arabiclistc}

\newcommand{\fcaption}[1]{
        \refstepcounter{figure}
        \setbox\@tempboxa = \hbox{\footnotesize Fig.~\thefigure. #1}
        \ifdim \wd\@tempboxa > 5in
           {\begin{center}
        \parbox{5in}{\footnotesize\smalllineskip Fig.~\thefigure. #1}
            \end{center}}
        \else
             {\begin{center}
             {\footnotesize Fig.~\thefigure. #1}
              \end{center}}
        \fi}
 
\newcommand{\tcaption}[1]{
        \refstepcounter{table}
        \setbox\@tempboxa = \hbox{\footnotesize Table~\thetable. #1}
        \ifdim \wd\@tempboxa > 5in
           {\begin{center}
        \parbox{5in}{\footnotesize\smalllineskip Table~\thetable. #1}
            \end{center}}
        \else
             {\begin{center}
             {\footnotesize Table~\thetable. #1}
              \end{center}}
        \fi}
 
\def\@citex[#1]#2{\if@filesw\immediate\write\@auxout
        {\string\citation{#2}}\fi
\def\@citea{}\@cite{\@for\@citeb:=#2\do
        {\@citea\def\@citea{,}\@ifundefined
        {b@\@citeb}{{\bf ?}\@warning
        {Citation `\@citeb' on page \thepage \space undefined}}
        {\csname b@\@citeb\endcsname}}}{#1}}
 
\newif\if@cghi
\def\cite{\@cghitrue\@ifnextchar [{\@tempswatrue
        \@citex}{\@tempswafalse\@citex[]}}
\def\citelow{\@cghifalse\@ifnextchar [{\@tempswatrue
        \@citex}{\@tempswafalse\@citex[]}}
\def\@cite#1#2{{$\null^{#1}$\if@tempswa\typeout
        {IJCGA warning: optional citation argument
        ignored: `#2'} \fi}}

\def\pmb#1{\setbox0=\hbox{#1}
        \kern-.025em\copy0\kern-\wd0
        \kern.05em\copy0\kern-\wd0
        \kern-.025em\raise.0433em\box0}

\def\fnt#1#2{\footnotetext{\kern-.3em
        {$^{\mbox{\scriptsize #1}}$}{#2}}}
 
\def\fpage#1{\begingroup
\voffset=.3in
\thispagestyle{empty}\begin{table}[b]\centerline{\footnotesize #1}
        \end{table}\endgroup}
 
\font\tenrm=cmr10
\font\tenit=cmti10
\font\tenbf=cmbx10
\font\bfit=cmbxti10 at 10pt
\font\ninerm=cmr9

\font\eightrm=cmr8

\def\qed{\hbox{${\vcenter{\vbox{                        
   \hrule height 0.4pt\hbox{\vrule width 0.4pt height 6pt
   \kern5pt\vrule width 0.4pt}\hrule height 0.4pt}}}$}}
 
\renewcommand{\thefootnote}{\fnsymbol{footnote}}        
 
\def\bsc{{\sc a\kern-6.4pt\sc a\kern-6.4pt\sc a}}       
\def\bflatex{\bf L\kern-.30em\raise.3ex\hbox{\bsc}\kern-.14em
T\kern-.1667em\lower.7ex\hbox{E}\kern-.125em X}
 
\catcode`@=11
\def\citer{\@ifnextchar [{\@tempswatrue\@citexr}{\@tempswafalse\@citexr[]}}

\def\@citexr[#1]#2{\if@filesw\immediate\write\@auxout
        {\string\citation{#2}}\fi
\def\@citea{}\@cite{\@for\@citeb:=#2\do
        {\@citea\def\@citea{-}\@ifundefined
        {b@\@citeb}{{\bf ?}\@warning
        {Citation `\@citeb' on page \thepage \space undefined}}
        {\csname b@\@citeb\endcsname}}}{#1}}
\catcode`@=12

\newcommand{\gsim}{\raisebox{-0.13cm}{~\shortstack{$>$ \\[-0.07cm] $\sim$}}~}

\newcommand{\SM}{\mbox{${\cal SM}$}}
\newcommand{\MSSM}{\mbox{${\cal MSSM}$}}
\newcommand{\SUSY}{\mbox{${\cal SUSY}$}}
\newcommand{\CP}{\mbox{${\cal CP}$}}
\newcommand{\HIGLU}{HIGLU}
\newcommand{\HDECAY}{HDECAY}

\newcommand{\tb}{\mbox{tg$\beta$}}

\newcommand{\nn}{\noindent}

\begin{document}
 
\begin{titlepage}

\begin{flushright}
CERN--TH/96--285 \\
hep--ph/9610350
\end{flushright}

\vspace{1cm}

\begin{center}

{\large\sc \HIGLU~and \HDECAY: Programs for \\
           Higgs Boson Production at the LHC and \\
           Higgs Boson Decay Widths\footnote{Talk given at the AIHENP'96
           Workshop, Lausanne, September 1996.}}

\end{center}

\vspace{0.5cm}

\begin{center}

{\large \sc Michael Spira}

\vspace{0.5cm}

CERN, TH Division, CH--1211 Geneva 23, Switzerland

\end{center}

\vspace{1cm}

\begin{abstract}
\nn
The total cross section for Higgs boson production via the dominant gluon
fusion mechanism including NLO [two-loop] QCD corrections can be
calculated numerically with the program \HIGLU. The QCD corrections are
included for arbitrary Higgs and quark masses and
increase the cross section at the LHC by up to a factor of 2. The source code
HIGLU provides the evaluation of the production of the Standard Model [\SM]
Higgs boson as well as the neutral Higgs bosons of the minimal
supersymmetric extension [\MSSM]. \\
\indent
The program \HDECAY~determines the decay widths and branching ratios of the
Higgs bosons within the \SM~and the \MSSM, including the dominant higher-order
corrections. The latter are dominated by QCD corrections and
two-loop corrections to the couplings and Higgs
masses of the \MSSM. The program includes all decay modes with branching
ratios larger than $10^{-4}$. Moreover, below-threshold decays
with off-shell top quarks, gauge and Higgs bosons are implemented.
In addition the program is able to calculate the branching
ratios of the \MSSM~Higgs bosons into supersymmetric particles, which can
be dominant.
\end{abstract}

\vspace{0.5cm}

\begin{flushleft}
CERN--TH/96--285 \\
hep--ph/9610350 \\
October 1996
\end{flushleft}

\end{titlepage}

\normalsize\textlineskip
\thispagestyle{empty}
\setcounter{page}{1}
 
\copyrightheading{}                     
 
\vspace*{0.88truein}
 
\addtocounter{footnote}{-1}
\fpage{1}
\centerline{\bf
\HIGLU~and \HDECAY: Programs for Higgs Boson Production}
\vspace*{0.035truein}
\centerline{\bf
at the LHC and Higgs Boson Decay Widths}
\vspace*{0.37truein}
\centerline{\footnotesize
\large \sc Michael Spira\footnote{e--mail address: spira@cern.ch}}
\vspace*{0.015truein}
\centerline{\footnotesize\it CERN Theory Division}
\baselineskip=10pt
\centerline{\footnotesize\it CH-1211 Geneva 23, Switzerland}
\vspace*{0.225truein}
 
\vspace*{0.21truein}
\abstracts{
The total cross section for Higgs boson production via the dominant gluon
fusion mechanism including NLO [two-loop] QCD corrections can be
calculated numerically with the program \HIGLU. The QCD corrections are
included for arbitrary Higgs and quark masses and
increase the cross section at the LHC by up to a factor of 2. The source code
HIGLU provides the evaluation of the production of the Standard Model [\SM]
Higgs boson as well as the neutral Higgs bosons of the minimal
supersymmetric extension [\MSSM]. \\
\indent
The program \HDECAY~determines the decay widths and branching ratios of the
Higgs bosons within the \SM~and the \MSSM, including the dominant higher-order
corrections. The latter are dominated by QCD corrections and
two-loop corrections to the couplings and Higgs
masses of the \MSSM. The program includes all decay modes with branching
ratios larger than $10^{-4}$. Moreover, below-threshold decays
with off-shell top quarks, gauge and Higgs bosons are implemented.
In addition the program is able to calculate the branching
ratios of the \MSSM~Higgs bosons into supersymmetric particles, which can
be dominant.}{}{}
 
 
 
\vspace*{1pt}\textlineskip      
\textheight=7.8truein
\setcounter{footnote}{0}
\renewcommand{\thefootnote}{\alph{footnote}}
 
\section{Introduction}
\noindent
At the future CERN hadron collider, the LHC, the Higgs boson will be produced
primarily via the gluon fusion mechanism for the entire relevant Higgs
mass range within the Standard Model [\SM] \cite{glufus} as well as its
minimal supersymmetric extension [\MSSM] \cite{glufumssm}.
The two-loop QCD corrections to the production cross
sections of scalar [\CP-even] as well as pseudoscalar [\CP-odd]
Higgs bosons have been calculated in \citer{QCDcorr,QCDcorr0} for heavy squarks.
The QCD corrections are significant
for the theoretical prediction of the cross sections leading to an
increase by up to a factor of 2 compared with the lowest-order
results.
Recently, the leading part of the pure QCD corrections to the squark loop
contributions to \MSSM~Higgs boson production have also been evaluated by
means of low-energy theorems in the limit of heavy gluinos
\cite{squark}. The squark loops are sizeable for squark masses below
$\sim 400$ GeV. The corrections are of the same magnitude as the corrections
to the quark loops so that the $K$ factors are only slightly changed by
the inclusion of the squark loops.

The program \HIGLU\footnote{Comments or suggestions are
welcome and should be sent to spira@cern.ch.}~\ provides the calculation of the
total Higgs production cross sections including next-to-leading order
QCD corrections to the heavy quark loops \cite{higlu}. Various input parameters
can be chosen from an input file, including a flag specifying the model.
Possible options are the Standard Model, its minimal supersymmetric extension,
and a general Higgs model by initializing the Higgs Yukawa
couplings appropriately. The program includes
the contribution of the top and bottom quarks in
the loops that generate the Higgs couplings to gluons. Within the Standard
Model, as well as in most of the parameter space of the
\MSSM, these contributions provide an excellent
approximation for all practical cases. Moreover, the program allows the
calculation of the
decay widths of Higgs bosons into gluons including next-to-leading
order QCD corrections. The gluonic decay mode plays a significant
r\^ole in the intermediate mass range at future $e^+e^-$ colliders
\cite{QCDcorr,QCDlim}.

The search for Higgs bosons at the LHC mainly proceeds via looking for
$ZZ$, $W^+W^-$ and $\gamma\gamma$ final states as decay products of the
\SM~Higgs particle and in addition for Higgs boson pairs in the case of the
\MSSM~Higgs bosons. Thus it is of vital importance to have reliable
predictions for the branching ratios of the Higgs bosons in both models.
These are determined by the program \HDECAY~including all relevant
higher-order corrections, which are dominated by QCD corrections and the
two-loop corrections to the \MSSM~Higgs masses and couplings arising from the
large top mass. \HDECAY~\cite{hdecay} calculates all decay modes with
branching ratio larger
than $10^{-4}$ and includes below-threshold decays involving off-shell
top quarks, gauge and Higgs bosons in the final states \cite{below}. Moreover,
all decay modes of the \MSSM~Higgs bosons into \SUSY~particles are contained
in the program, which can be dominant for Higgs masses below the
$t\bar t$-threshold.

The source codes of the programs are written in FORTRAN. They have been
tested on computers  running under different operating systems. The numerical
phase space integration of \HIGLU~is performed by using the VEGAS package
\cite{VEGAS} for integrals of dimension up to 3. Parton distributions can
be attached to \HIGLU~in any desirable way by adjusting the
corresponding subroutine.
As standard parametrization, the program contains the GRV sets \cite{GRV}.

\section{Processes}
\subsection{$pp \to \Phi^0 + X$}
The hadron cross section of Higgs boson\footnote{The scalar [$\cal CP$-even]
Higgs particles will generically be denoted by $\cal H$, the pseudoscalar
[$\cal CP$-odd] by $A$ and all the neutral Higgs bosons by $\Phi^0$. The
\MSSM~Higgs particles, including the charged ones, will be denoted by
$\Phi$.}~\ production via gluon fusion
$gg\to \Phi^0$ $(\Phi^0 = {\cal H},A)$ including
[two-loop] QCD corrections, can be cast into the form
\begin{equation}
\sigma (pp \to \Phi^0 + X) = \sigma^{\Phi^0}_{LO}
+ \Delta \sigma^{\Phi^0}_{virt} + \Delta \sigma^{\Phi^0}_{gg}
+ \Delta \sigma^{\Phi^0}_{gq} + \Delta \sigma^{\Phi^0}_{q\bar q} \, .
\label{eq:glufus}
\end{equation}
The lowest-order cross sections $\sigma_{LO}^{\Phi^0}$ can be found in
Refs.\citer{glufus,QCDcorr0}. The term $\Delta
\sigma^{\Phi^0}_{virt}$ parametrizes the infrared regularized virtual two-loop
corrections and the terms $\Delta \sigma^{\Phi^0}_{ij}~\ (i,j = g,q, \bar q)$
the individual collinear regularized real one-loop corrections corresponding
to the subprocesses
\begin{equation}
gg \to \Phi^0 g, \hspace{1cm} gq \to \Phi^0 q, \hspace{1cm} q\bar q \to
\Phi^0 g \, .
\label{realcorr}
\end{equation}
Their expressions can be found in Refs.\citer{QCDcorr,QCDcorr0}.

The calculation of the [two-loop] virtual corrections requires the evaluation of
massive two-loop three-point functions depending on one free parameter, the
ratio of the Higgs and heavy quark masses. The five-dimensional 
Feynman integrals have been analytically reduced to one-dimensional integrals;
they are numerically integrated by an adaptive Romberg integration
\cite{DECADRE}. The regularization of UV, IR and collinear divergences is
performed in $n=4-2\epsilon$ dimensions. The UV singularities are absorbed in
the strong coupling $\alpha_s$ and the on-shell quark mass $m_Q$.
The infrared singularities of
the virtual corrections cancel against the corresponding singularities
of the real corrections of the subprocesses given in Eq.~(\ref{realcorr}).
The left-over collinear poles are mapped into the renormalized NLO parton
densities.
The renormalization and factorization schemes have been chosen as the
$\overline{MS}$ scheme.
 
The program \HIGLU~calculates the five terms in Eq.~(\ref{eq:glufus})
contributing to the total cross section separately, as well as their sum, for
all kinds of neutral Higgs bosons $\Phi^0$.

\subsection{$\Phi^0 \to gg$}
The decay widths of Higgs bosons $\Phi^0$ into gluons up to next-to-leading
order are given by
\begin{eqnarray}
\Gamma (\Phi^0 \to gg(g), gq\bar q) & = & \Gamma_{LO} (\Phi^0 \to gg) \left[ 1 +
E_{\Phi^0} \frac{\alpha_s}{\pi} \right] \nonumber \\ \nonumber \\
E_{\Phi^0} & = & E_{virt}^{\Phi^0} + E_{ggg}^{\Phi^0} +
N_F E_{gq\bar q}^{\Phi^0} \, .
\end{eqnarray}
The coefficient $E_{virt}^{\Phi^0}$ denotes the infrared-regularized virtual
two-loop
corrections, $E_{ggg}^{\Phi^0}$ and $E_{gq\bar q}^{\Phi^0}$ the infrared and 
collinear regularized
real one-loop corrections.  The analytical formulae of these contributions
can be obtained from the gluon fusion process by crossing and are given
in Refs.\cite{QCDcorr,QCDlim,inami}.  The parameter $N_F$ fixes the number
of light external
flavors produced in the decay $\Phi^0\to gq\bar q$, which has to be identified
with the number of flavors contributing to the QCD $\beta$ function in order
to map large logarithms into the running strong coupling
$\alpha_s (\mu^2)$.
This issue is important, if bottom and charm quarks will not be included as
external flavors in the gluonic decay mode, but added to the $b\bar b$ and
$c\bar c$ decay channels: in order to avoid large
logarithms in the perturbative expansion, the strong coupling has to be
evolved with three light flavors instead of five at scales above the
heavy-quark thresholds.
The program \HIGLU~evaluates this
decay mode, including the full mass dependence. The program \HDECAY~on the
other hand only contains the leading part in the heavy-quark mass limit,
which is a reliable approximation in all relevant cases.
The gluonic branching ratios reach values of about 10--20\% in the intermediate
mass ranges and are thus significant for phenomenological analyses.

\subsection{$\Phi^0 \to Q\bar Q$, $H^+ \to U\bar D$}
The expressions for decay widths of Higgs bosons into heavy-quark pairs
can be found in Ref.~\cite{HQQ}.
In order to absorb large logarithms in the limit $M_\Phi \gg m_Q$, one has to
introduce the running $\overline{MS}$ mass for the heavy quarks in the final
states, which has to be evaluated at the scale of the corresponding Higgs
boson mass $M_\phi$. This definition leads to a well-behaved perturbative
expansion in the case of charm and bottom quarks above the thresholds.
The running $\overline{MS}$ mass is calculated from the value of the pole
mass by means of the relation given in Ref.\cite{polemass}. A flag is provided
for the choice of the NLO or NNLO formula of this relation, which is important
in particular for the charm-quark mass, because of a bad convergence
of the perturbative expansion at scales of the order of the charm mass.
The $\overline{MS}$ mass characterizes the preferred input
parameter, because it can be obtained directly from fits to QCD sum rules
\cite{Narison}.

In the limit $M_\Phi \gg m_Q$ the QCD corrections are known up to NNLO. At
the threshold, mass effects are important. Thus we have included the full
massive NLO QCD corrections in the program \HDECAY, which are smoothly
interpolated with the NNLO expressions in the large-Higgs mass range.
The QCD corrections reduce the decay widths significantly.
Electroweak corrections are small \cite{helw} and hence neglected in the
program. The $b\bar b$ decay mode dominates in the intermediate neutral
Higgs mass range,
if decays into \SUSY~particles are kinematically forbidden,
except for the heavy scalar \MSSM~ Higgs particle $H$ for small \tb.
The decays into $c\bar c$ and
$\tau^+\tau^-$ pairs can reach branching ratios of about 10\%.
Above threshold the $t\bar t$ decay modes are
dominating in the \MSSM. Charged \MSSM~Higgs bosons predominantly decay into
$\tau\nu$ and $tb$ pairs, if \SUSY~particle decays are impossible. For top
quarks in the final states, \HDECAY~ also takes into account below-threshold
decays into off-shell top quarks \cite{below}, which are smoothly
interpolated with the on-shell regime.

\subsection{$\Phi^0 \to \gamma\gamma$}
The photonic decay widths of the neutral Higgs particles are built-up by
charged-particle loop contributions from fermions and $W$ bosons in the
\SM~and in addition from charged Higgs particles, charginos and sfermions in the
\MSSM. The expressions can be found in Ref.\cite{QCDcorr}. The QCD corrections
are moderate \cite{QCDcorr} and thus neglected in the program \HDECAY.
The photonic branching ratios reach values of about $10^{-3}$. These decay modes
are important for the Higgs search at the LHC \cite{LHCgam}.

\subsection{${\cal H} \to WW, ZZ$}
The expressions for the scalar decay widths into $W$ and $Z$ boson pairs
can be found in Ref.\cite{WZH}. They are the dominant  decay modes of the
\SM~Higgs boson for Higgs masses $M_H\gsim 140$ GeV. In the \MSSM~they
are only important for small \tb. The program HDECAY~includes below-threshold
decays into off-shell $W$ and $Z$ bosons \cite{below}, which are interpolated
with the expressions of the on-shell range. Electroweak corrections are small
\cite{helw} in the relevant Higgs mass ranges and neglected in \HDECAY.

\subsection{Decays into Higgs particles}
The decay widths of the \MSSM~Higgs bosons into lighter Higgs particles are
given in Ref.\cite{WZH}. The decay mode $H\to hh$ is dominant below the
$t\bar t$-threshold for small \tb. In parts of the mass ranges, the decay
modes $A\to Zh$ and $H^\pm\to W^\pm h$ are dominating. \HDECAY~also includes
below-threshold decays into off-shell $Z$ and Higgs bosons \cite{below} and
interpolates
these contributions with the on-shell regime. The dominant corrections are
contained in the two-loop-corrected \MSSM~couplings.

\subsection{Decays into \SUSY~particles}
The partial decay widths into neutralinos, charginos and sfermions are given
in Ref.\cite{SUSY}. Their branching ratios can amount up to about 100\%
below the top threshold for small \tb~and remain significant above \cite{SUSY}.
\HDECAY~evaluates all these partial decay widths within the \MSSM, in which
they are completely fixed by two \MSSM~parameters, $\mu$ and $M_2$.

\section{Input and Ouput}
All masses, couplings and \MSSM~parameters can be defined in separate
input files for both programs \HIGLU~and \HDECAY. Moreover, there are a few
flags to allow a choice of the kind of process, and of the conventions
for the parameters involved in the different processes. Thus the two programs
provide a very flexible and convenient usage, fitting to all options of
phenomenological relevance.

The program \HIGLU~writes all chosen input parameters as well as the results
obtained by VEGAS integrations including the individual parts of the QCD
corrections to the file unit 99.

\HDECAY~produces many output files containing the results for the Higgs 
branching ratios and total decay widths in table format. These can easily
be read by other FORTRAN programs, which need the values for the Higgs
branching ratios and decay widths.

\nonumsection{Acknowledgements}
\noindent
I would like to thank A.~Djouadi, P.M.~Zerwas, J.~Kalinowski and D.~Graudenz
for their fruitful collaboration. Special thanks go to A.~Djouadi for a
critical reading of the manuscript and useful discussions.
 
\nonumsection{References}

\end{document}